\documentclass[a4paper, 12pt]{article}

\baselineskip=\normalbaselineskip

\usepackage[top=21truemm,bottom=30truemm,left=23truemm,right=23truemm]{geometry}
\usepackage{setspace}

\usepackage{mathrsfs,amsbsy,amssymb,latexsym,amsfonts,amsmath,amsthm}
\usepackage{graphicx,color}
\usepackage[subrefformat=parens]{subcaption}
\usepackage{braket}

\allowdisplaybreaks[2]
\renewcommand{\thefootnote}{\arabic{footnote}}
\makeatletter

\long\def\affiliation#1{\long\gdef\@affiliation{#1}}

\catcode`\@=11
\@addtoreset{equation}{section}

\def\@seccntformat#1{\csname the#1\endcsname.~~}
\makeatother

\newcommand{\nn}{\nonumber}
\newcommand{\rmd}{{\mathrm d}}
\DeclareMathOperator{\tr}{tr}
\begin{document}

\begin{titlepage}

\setcounter{page}{0}

\renewcommand{\thefootnote}{\fnsymbol{footnote}}

\begin{flushright}
OU-HET 903 
\end{flushright}

\vskip 1.35cm

\begin{center}
{\Large \bf 
Entanglement entropy for free scalar fields in AdS
}

\vskip 1.2cm 

{\normalsize
Sotaro Sugishita\footnote{sugishita(at)het.phys.sci.osaka-u.ac.jp} 
}

\vskip 0.3cm
{\it Department of Physics, Osaka University, Toyonaka, Osaka, 560-0043, Japan}

\end{center}

\vspace{8mm}

\centerline{{\bf Abstract}}

We compute entanglement entropy for free massive scalar fields in anti-de Sitter (AdS) space. 
The entangling surface is a minimal surface whose boundary is a sphere at the boundary of AdS. 
The entropy can be evaluated 
from the thermal free energy of the fields on a topological black hole 
by using the replica method.  
In odd-dimensional AdS, 
exact expressions of 
the R\'enyi entropy $S_n$ are obtained for arbitrary $n$.  
We also evaluate
1-loop corrections coming from the scalar fields to holographic entanglement entropy.  
Applying the results, we compute the leading difference of entanglement entropy 
between two holographic CFTs 
related by a renormalization group flow triggered by a double trace deformation.   
The difference is proportional to 
the shift of a central charge under the flow.

\end{titlepage}
\newpage

\tableofcontents
\vskip 1.2cm

\section{Introduction}

The study of 
entanglement entropy in quantum field theories began to give a microscopic 
explanation of the black hole entropy
\cite{Bombelli:1986rw,Srednicki:1993im,Callan:1994py,
Kabat:1994vj,Kabat:1995eq}. 
The area-law of the entanglement entropy of a region and its complement, 
which is also called geometric entropy, 
actually resembles the Bekenstein-Hawking entropy 
\cite{Bekenstein:1972tm,Bekenstein:1973ur,Bekenstein:1974ax,Hawking:1974sw}. 
Entanglement entropy 
is expected to be related to degrees of freedom in the system.  
For example, we can analytically calculate entanglement entropy for a single interval in (1+1)-dimensional conformal field theory 
\cite{Holzhey:1994we, Calabrese:2004eu,Calabrese:2009qy}, 
and it is proportional to the central charge of the CFT.  
However, it is generally difficult to directly compute 
entanglement entropy in higher dimensional CFTs and non-conformal field theories (
see, e.g.,  
\cite{Casini:2009sr,Casini:2010kt,Klebanov:2011uf,Banerjee:2015tia} 
where entanglement entropy in free theories is evaluated).  
In order to know general properties of entanglement entropy, 
we should investigate examples where entanglement entropy is computed analytically.  
A natural extension is to consider QFTs in curved backgrounds   
such as black hole backgrounds  
\cite{Solodukhin:2011gn},  
de Sitter space \cite{Maldacena:2012xp} 
and 
anti-de Sitter space (AdS) \cite{Miyagawa:2015sql}. 

\begin{figure}
\vspace{3mm}
\begin{center}
\includegraphics[height=4cm]{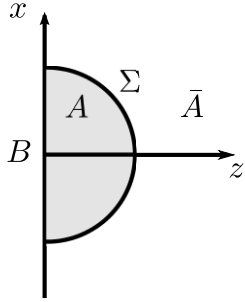}
\begin{quote}
\caption{Minimal surface $\Sigma$ corresponding to 
the region $B$ in the boundary. 
The $z$-direction denotes the bulk direction, and 
$x$ are the spatial coordinates of the boundary. 
The region surrounded by $B$ and $\Sigma$ is denoted by $A$. 
}
\label{fig:min}
\end{quote}
\end{center}
\end{figure}

In this paper, 
we compute entanglement entropy for free massive scalar fields in AdS. 
Applying the result, we can evaluate quantum corrections of holographic entanglement entropy \cite{Ryu:2006bv} as in \cite{Miyagawa:2015sql}. 
The holographic entanglement entropy formula is proposed,    
in the context of the AdS/CFT correspondence \cite{Maldacena:1997re},  
as a simple formula to compute entanglement entropy of a CFT 
with a gravity dual 
(see also \cite{Ryu:2006ef, Nishioka:2009un}).   
The formula states that 
the entanglement entropy of region $B$ in a CFT is, 
like the Bekenstein-Hawking entropy formula,  
proportional to the minimal area of a bulk surface $\Sigma$ that ends on the boundary of $B$ 
(see Fig~\ref{fig:min}), 
\begin{align}
S_{cl} (B) = \frac{\text{area}(\Sigma)}{4 G_N}, 
\label{RT} 
\end{align}
where $G_N$ is bulk Newton's constant. 
This formula is valid at the classical level (in the bulk).\footnote{
If we consider higher derivative gravity, the formula is replaced by the classical Wald-like entropy formula (see, e.g., \cite{Hung:2011xb,Dong:2013qoa,Camps:2013zua}).} 
If the dual CFT is a large $N$ theory, 
the contribution of \eqref{RT} corresponds to order $N^2$. 
In order to include the $1/N$ corrections in the CFT side, 
we need to consider quantum corrections to eq.~\eqref{RT}.    
In other words, the formula \eqref{RT} 
is the leading term in the $G_N$ expansion, which is order $G_N^{-1}$. 
  
Faulkner, Lewkowycz and Maldacena (FLM) \cite{Faulkner:2013ana} propose that
the $\mathcal{O}(G_N^0)$ correction to the holographic entanglement entropy 
consists as follows 
\begin{align}
S_q(B) = S^\text{bulk}(A) 
+ \delta\frac{\text{area}(\Sigma)}{4 G_N} 
+\delta S_\text{Wald} + S_{c.t.} \,.  
\label{FLM}
\end{align} 
The first term $S^\text{bulk}(A)$ represents 
the entanglement entropy of quantum fields between a region $A$ and its complement $\bar{A}$ in the bulk.\footnote{
Note that it has been discussed in the context of black hole entropy that 
entanglement entropy of matter fields can be interpreted as the quantum correction to  
the Bekenstein-Hawking entropy (see, e.g., a review paper \cite{Solodukhin:2011gn}). } 
Here, $A$ is the region surrounded by $B$ and $\Sigma$ 
as in Fig.~\ref{fig:min}. 
The second term is the shift of the minimal area due to the change of the background because of quantum expectation values of matter fields. 
The term $\delta S_\text{Wald}$ denotes Wald-like entropy contributions arising from the expectation values of quantum fields. 
The last term $S_{c.t.}$ is introduced as the counter terms to cancel 
the bulk UV divergences.

In \cite{Miyagawa:2015sql}, Miyagawa, Shiba and Takayanagi investigate an example where the quantum corrections \eqref{FLM} give the leading contributions.\footnote{
See also \cite{Agon:2015ftl} where quantum corrections of holographic mutual information is computed, 
which is another example that quantum corrections are the leading contributions.} 
They consider a gravity dual of a CFT perturbed by a relevant double trace deformation \cite{Witten:2001ua,Gubser:2002zh,Gubser:2002vv} 
and 
study the change of holographic entanglement entropy under a 
flow produced by the double trace deformation. 
In the gravity side, 
there is a massive scalar field dual to 
a single trace operator $\mathcal{O}$. 
The dimension of operator $\mathcal{O}$ is related  to the  
mass $m$ of the scalar field \cite{Gubser:1998bc,Witten:1998qj} 
as 
\begin{align}
\Delta_\pm = \frac{d-1}{2} \pm \nu, \quad\quad 
\nu = \sqrt{m^2 \,\ell_\text{AdS}^2  + \biggl(\frac{d-1}{2}\biggr)^2} \,,
\label{def_nu}  
\end{align}
where $d$ denotes the dimensions of AdS space and 
$\ell_\text{AdS}$ is the radius of AdS. 
When the mass of scalar field is in a certain range 
\begin{align}
-\biggl(\frac{d-1}{2}\biggr)^2 < m^2 \ell_\text{AdS}^2 < 
-\biggl(\frac{d-1}{2}\biggr)^2 +1 \,,
\end{align}
that is, $0<\nu<1$, 
both dimensions $\Delta_\pm$ satisfy the unitarity bound 
$\Delta_\pm>(d-3)/2$ of $(d-1)$-dimensional CFT,   
and two corresponding boundary conditions of the scalar field 
in AdS are allowed \cite{Klebanov:1999tb}. 
One is the Dirichlet boundary condition corresponding to $\Delta_+$, 
and the other is the Neumann boundary condition 
corresponding to $\Delta_-$.  
If we start from a CFT, (we call CFT$^{(N)}$), where $\mathcal{O}$ has the dimension $\Delta_-$ 
and add a double trace deformation $\mathcal{O}^2$, 
which is relevant, 
the theory flows to another CFT (we call CFT$^{(D)}$) 
where the dimension of $\mathcal{O}$ is $\Delta_+$. 
In the dual gravity side, the difference of two theories is the boundary conditions of the scalar field. 
Thus, the leading contributions of holographic entanglement entropy 
\eqref{RT} are
the same for both theories. 
In addition, contributions from other fields in the bulk are not affected 
by the difference of the scalar boundary conditions at the 1-loop level. 
Therefore, if we consider the difference of entanglement entropy between 
CFT$^{(N)}$ and CFT$^{(D)}$, 
the leading difference comes from 1-loop contributions 
\eqref{FLM} of the scalar field. 

The subregion $B$ in CFT is taken to be a half space 
in \cite{Miyagawa:2015sql}. 
In the present paper, we take the subregion $B$ 
as a ball with radius $r_0$. 
In fact, if the subregion is a ball, the universal part of entanglement 
entropy is given by \cite{Casini:2011kv}
\begin{align}
S^{univ}(B) = 
\left\{ \begin{array}{ll}
(-1)^\frac{d-3}{2} \,4\, a_{d-1}^\ast \log \frac{r_0}{\epsilon}  
 & (d-1:\text{even}) \\
(-1)^\frac{d-2}{2} \, 2 \pi \, a_{d-1}^\ast
 & (d-1:\text{odd}) \\
\end{array} \right. \,, 
\label{suniv_anomaly}
\end{align} 
where $\epsilon$ is a boundary UV cutoff, 
and $a_{d-1}^\ast$ agrees with 
the A-type trace anomaly $a$ in the case where 
$d-1$ is even (see also \cite{Myers:2010tj}). 
Since the shift of the central charge $a$ under the double trace deformation is 
computed at the leading order without AdS/CFT in \cite{Gubser:2002vv},\footnote{
It is also computed in \cite{Gubser:2002zh} holographically.} 
we can test the FLM proposal \eqref{FLM} by comparing the change of entanglement entropy with the result in \cite{Gubser:2002vv}. 
We will evaluate, in odd-dimensional AdS$_{d}$ $(3\leq d\leq 11)$, 
all terms in \eqref{FLM} except for $S_{c.t.}$, 
where we assume that $S_{c.t.}$ just cancels the bulk UV divergences of the other terms. 
The result is consistent with that expected from \eqref{suniv_anomaly}.

We also give explicit expressions of the R\'enyi 
entanglement entropy for free massive scalar fields in 
odd-dimensional AdS$_{d}$,  
by a purely field theoretic computation. 
We hope that our results serve as an example of 
the R\'enyi entropy for non-conformal theories.

This paper is organized as follows: 
In section \ref{sec_method} 
we summarize a method for computing entanglement entropy in AdS. 
In section \ref{sec_eeads}, 
we compute entanglement entropy (and the R\'enyi entropy) 
for free massive scalar fields 
in odd-dimensional AdS using the heat kernels. 
In section \ref{sec_1loop}, we evaluate 
1-loop corrections of holographic entanglement entropy 
and find that 
the change of entanglement entropy under an RG flow 
by a double trace deformation is proportional to the shift of the A-type 
central charge. 
In section \ref{sec_sum}, we summarize our results and 
give some discussion.  

\section{Method for computing entanglement entropy in AdS}
\label{sec_method}

In this paper, we use the replica method to compute 
entanglement entropy in AdS, 
which is reviewed in subsection \ref{subsec_replica}. 
Using the replica method, the R\'enyi entropy can be computed from 
a free energy on a replicated space. 
We will see that the free energy is given by a thermal free energy on the topological black hole 
in subsection \ref{subsec_topo}. 
The fact also enables us to compute the modular Hamiltonian of the bulk fields. 
In subsection \ref{subsec_arealaw}, we will confirm that 
the leading divergence of R\'enyi entropy satisfies the area-law 
for general dimensions and general mass.

\subsection{Replica method} 
\label{subsec_replica}

In this subsection, we review the replica method (see, e.g., 
\cite{Callan:1994py, Calabrese:2004eu, Calabrese:2009qy}).   

We consider a theory on $d$-dimensional AdS space, 
and  compute entanglement entropy of a region $A$ for the ground state. 
The total density matrix of the ground state $\rho_\text{tot}$ can be represented as a path integral  
\begin{align}
\Bra{\phi_1(\vec{x})}\rho_\text{tot} \Ket{\phi_2(\vec{x})} =\frac{1}{Z_1} 
\int^{t_E =\infty}_{t_E=-\infty} \!\!\!\! \mathcal{D} \phi(t_E, \vec{x}) \, \delta(\phi(0_-,\vec{x})-\phi_1(\vec{x})) \, \delta(\phi(0_+,\vec{x})-\phi_2(\vec{x})) \, e^{-S_E} \,, 
\end{align}
where $S_E$ is the Euclidean action and $Z_1$ is the partition function:  
\begin{align}
Z_1 = \int^{t_E =\infty}_{t_E=-\infty} \!\!\!\! \mathcal{D} \phi(t_E, \vec{x}) \, e^{-S_E} \,. 
\end{align}
Using the path integral representation, the reduced density matrix on the region $A$ is given by 
\begin{align}
&\Bra{\phi_1^A(\vec{x})}\rho_A\Ket{\phi_2^A(\vec{x})} \nn\\
&=\frac{1}{Z_1} 
\int^{t_E =\infty}_{t_E=-\infty} \!\!\!\! \mathcal{D} \phi(t_E, \vec{x}) \,
\prod_{\vec{x}\in A} \delta(\phi(0_-,\vec{x})-\phi_1^A(\vec{x})) \, \delta(\phi(0_+,\vec{x})-\phi_2^A(\vec{x})) \, e^{-S_E} \,.  
\end{align}
Thus, we have 
\begin{align}
\tr_A\rho_A^n = \frac{Z_n}{Z_1^n} \,,
\end{align} 
where $Z_n$ represents a path integral on $n$-sheeted covering space  
$\mathcal{M}_n$ which is obtained by sewing cyclically 
$n$ copies of the original Euclidean AdS space (EAdS) together along $A$. 
The R\'enyi entanglement entropy $S_n$ is then represented as 
\begin{align}
S_n \equiv \frac{\log \tr_A\rho_A^n}{1-n} 
=\frac{1}{1-n} (\log Z_n- n \log Z_1) \,. 
\end{align}  
If we obtain the analytic continuation of $S_n$ to $\mathrm{Re}\,n>1$, 
the (von Neumann) entanglement entropy $S_1$ is computed as 
\begin{align}
S_1 \equiv  -\tr_A \rho_A \log \rho_A = \lim_{n\to 1} S_n \,. 
\end{align}

\subsection{Coordinate transformations and topological black hole}
\label{subsec_topo}
In the Poincar\'e coordinates, 
the metric of AdS space is given by 
\begin{align}
\rmd s^2 = \ell_\text{AdS}^2 \frac{\rmd z^2-\rmd t^2 + \sum_{i=1}^{d-2}\rmd x_i^2}{z^2}\,,  
\label{poincare}
\end{align}
where $\ell_\text{AdS}$ is the radius of AdS$_d$. 
We also write the coordinates as 
\begin{align}
z=r \sin\theta, \, x_i=r \cos\theta\, \Omega_i
\label{polar_co}
\end{align}
with
\begin{align}
r>0,\, \quad 
\left\{ \begin{array}{ll}
0<\theta < \pi
 & (d=3) \\
0<\theta<\pi/2
 & (d\geq 4) \\
\end{array} \right. \,, 
\end{align}
where $\Omega_i$ denote coordinates of $(d-3)$-dimensional sphere. 
We consider the minimal surface $\Sigma$ corresponding to 
a ball region $B$ with radius $r_0$.  
One can find that the minimal surface $\Sigma$ is given by 
$t=0,\, r=r_0$ \cite{Ryu:2006bv}.  
We thus compute entanglement entropy between 
the inside region $A=\{t=0, r<r_0\}$ and its complement $\bar{A}=\{t=0, r>r_0\}$ for the ground state,   
using the replica method. 

In the dual CFT side, 
there is a conformal transformation 
\cite{Casini:2010kt,Casini:2011kv} 
that 
the causal development of spatial ball $B$ 
is mapped to $\mathbf{R}\times H^{d-2}$ where 
$H^{d-2}$ is $(d-2)$-dimensional hyperbolic space. 
Then the reduced density matrix on the ball for the vacuum state is 
mapped to a thermal state. 
Entanglement entropy for the ball region is thus 
the thermal entropy on the $H^{d-2}$. 
If the AdS/CFT corresponding is valid, 
the thermal entropy is equal to 
entropy for a topological black hole \cite{Casini:2011kv}.   
The classical contribution 
is given by the horizon entropy of 
the topological black hole. 
Matter fields on the topological black hole 
also contribute to the thermal entropy as the quantum corrections. 
This is the reason why the bulk entanglement entropy 
gives a quantum correction to holographic entanglement entropy.  
We will explicitly write the corresponding coordinate transformation 
in the bulk space 
such that region $A$ is mapped to the outside of the horizon 
in a topological black hole
and see the entanglement entropy of $A$ is equal to a thermal entropy 
on the black hole.

Using the coordinates \eqref{polar_co}, 
the metric of Euclidean AdS$_d$ is  written as 
\begin{align}
\rmd s^2 =  \frac{\rmd r^2 +\rmd t_E^2 + r^2 \rmd \theta^2  +r^2 \cos^2 \theta \, \rmd \Omega_{d-3}^2}{r^2 \sin^2\theta}\,, 
\label{eads_metric1}
\end{align}
where we set $\ell_\text{AdS}=1$.  

\begin{figure}
\vspace{3mm}
\begin{center}
\includegraphics[height=4cm]{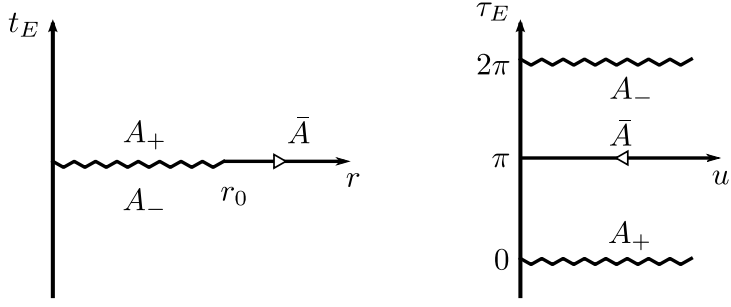}
\begin{quote}
\caption{Coordinate transformation from $(r,t_E)$ to $(u,\tau_E)$.  }
\label{fig:map}
\end{quote}
\end{center}
\end{figure}

We then transform the coordinates   
as in \cite{Casini:2011kv, Hung:2014npa}:  
\begin{align}
r=r_0 \frac{\sinh u}{\cosh u+ \cos \tau_E}\,, \quad 
t_E=r_0 \frac{\sin \tau_E}{\cosh u+ \cos \tau_E}\,. 
\label{co_tr}
\end{align}  
In the coordinates $(u, \tau_E, \theta, \Omega_i)$, 
(where $u>0$, $0\leq \tau_E <2\pi$), 
the metric \eqref{eads_metric1} takes the form 
\begin{align}
\rmd s^2 =
\frac{\rmd u^2 +\rmd \tau_E^2 + \sinh^2 u \, ( \rmd \theta^2  + \cos^2 \theta\,  \rmd \Omega_{d-3}^2)}{\sinh^2 u\,  \sin^2\theta}\,. 
\end{align}
Under the coordinate transformation, 
as shown in Fig.~\ref{fig:map}, 
neighborhoods $A_\pm =\{r<r_0,\, t_E=0\pm\}$ 
of region $A=\{r<r_0,\, t_E=0\}$   
are respectively mapped to $\{u>0, \tau_E=0+\}$ and $\{u>0, \tau_E=2\pi-\}$, 
and the complement $\bar{A} = \{r>r_0, \, t_E=0\}$ is mapped to 
$\{u>0, \tau_E=\pi\}$. 

We thus obtain the covering space $\mathcal{M}_n$ by extending the period of $\tau_E$ from $2\pi$ to $2\pi n$, 
noting that there is a translational symmetry in the $\tau_E$-direction.

We also introduce other coordinates $(\rho, \tau_E, v, \Omega_i)$ defined by 
\begin{align}
\rho&=\frac{1}{\sin\theta \, \sinh u}
=\frac{\sqrt{((r_0+r)^2+t_E^2)((r_0-r)^2+t_E^2)}}{2\,r_0 \, r \sin\theta}\,, 
\\
v&= \frac12 \log \frac{1-\tanh u \cos \theta}{1+\tanh u \cos \theta} 
= \frac12 \log \frac{r_0^2 +r^2 + t_E^2 -2 r_0\, r \cos \theta}
{r_0^2 +r^2 + t_E^2 + 2 r_0\, r \cos \theta} \,,  
\end{align}
where $\rho> 0$ and  $0<v<\infty$ (if $d=3$, $-\infty<v<\infty$).  
The metric is then given by 
\begin{align}
\rmd s^2 =
\frac{\rmd \rho^2}{1+\rho^2}+\rho^2 \rmd \tau_E^2
+(1+\rho^2)(\rmd v^2 + \sinh^2 v \, \rmd \Omega_{d-3}^2)\,. 
\label{ds_topobh}
\end{align}
This is a metric of a (Euclidean) topological black hole\footnote{
A radial coordinate $\tilde{\rho}=\sqrt{\rho^2+1}$ is often used. 
Then, the horizon is given by $\tilde{\rho}=1$.  } 
whose horizon is a hyperbolic space $H^{d-2}$
at $\tau_E=\rho=0$.  
As mentioned above, 
the covering space $\mathcal{M}_n$ has the period $\tau_E\sim \tau_E+2\pi n$.  
Therefore,  
the partition function 
$Z_n$  is the thermal partition function at temperature $1/(2\pi n)$ 
on the topological black hole.

We also comment that the entangling surface $\Sigma=\{t_E=0, r=r_0\}$ is 
mapped to surface $\{\tau_E=\rho=0\}$, i.e., the horizon. 
The area of the entangling surface, $\mathcal{A}^{(d-2)} (\Sigma)$,  
is thus the area of the horizon \cite{Casini:2011kv}:  
\begin{align}
\mathcal{A}^{(d-2)} (\Sigma) &= \int_{H^{d-2}} \rmd V_{d-2} 
=\Omega_{d-3} \int^\infty_{0} \rmd v \, \sinh^{d-3}\! v \nn\\
&=  \Omega_{d-3} \int^{\pi/2}_0 \rmd \theta \,\frac{\cos^{d-3}\theta }{\sin^{d-2}\theta}\,, 
\label{area}
\end{align}  
where $\Omega_{d-3}$ represents the area of $(d-3)$-sphere   
\begin{align}
\Omega_{d-3} = \frac{2 \pi^\frac{d-2}{2}}{\Gamma(\frac{d-2}{2})} \,. 
\end{align}
The area $\mathcal{A}^{(d-2)} (\Sigma)$
is a divergent quantity since hyperbolic space $H^{d-2}$ is non-compact. 
If we introduce a cutoff surface at $z=\epsilon$ 
in the Poincar\'e coordinates \eqref{poincare}, 
the area is given by 
\begin{align}
 \mathcal{A}^{(d-2)} (\Sigma) &= \Omega_{d-3}  \int^{r_0}_\epsilon \frac{\rmd z}{r_0} \frac{(1-z^2/r_0^2)^{\frac{d-4}{2}}}{(z/r_0)^{d-2}} 
\nn\\
&= \Omega_{d-3}  \int^{1}_{\epsilon/r_0} \rmd y \frac{(1-y^2)^{\frac{d-4}{2}}}{y^{d-2}} 
\,. 
\label{area_finite}
\end{align}

\subsection{Modular Hamiltonian} 
Here we briefly comment on the modular Hamiltonian 
of the bulk scalar field. 
If we have a density matrix $\hat{\rho}$, the corresponding modular Hamiltonian $\hat{H}$ is 
defined by $\hat{H}=-\log \hat{\rho}$. 
In the case that we have considered in the previous subsection, 
the reduced density matrix $\rho_A$ represents a thermal state with respect to  Hamiltonian $H_\tau$ corresponding to 
a Killing vector $\xi = \xi^\mu \partial_\mu=\partial_\tau$ on the topological black hole:   
\begin{align}
\rho_A = \frac{e^{-2\pi H_\tau}}{\tr_A  e^{-2\pi H_\tau}}
\end{align}
with 
\begin{align}
H_\tau= \int_{\tau=0} \! T_{\mu\nu} \xi^\mu \rmd S^\nu
= \int \!\rmd \rho \, (1+\rho^2)^{\frac{d-3}{2}} \rho^{-1}
\int_{H^{d-2}}\!\! \rmd V_{d-2}\, 
T_{\tau\tau} \,.  
\end{align}
The modular Hamiltonian $H_A$  is given by $2\pi H_\tau$  
up to a constant operator   
as 
\begin{align}
H_A = 2\pi H_\tau + (const.)\,. 
\end{align}

\subsection{Area-law in the bulk}
\label{subsec_arealaw}
Although the above argument can be applied to general quantum field theories, 
we consider a free scalar field on $d$-dimensional AdS space:  
\begin{align}
S=\int\rmd^d x \sqrt{-g} \, \frac12 \bigl(-\partial_\mu \phi \partial^\mu \phi  -(m_0^2 +\xi R)\phi^2 \bigr) \,. 
\end{align}
Since the Ricci scalar $R$ of AdS space is constant  
\begin{align}
R=-\frac{d(d-1)}{\ell_\text{AdS}^2}\,, 
\end{align}
we include the curvature coupling term in the mass term and write as 
\begin{align}
m^2=m_0^2 -\xi \frac{d(d-1)}{\ell_\text{AdS}^2}\,. 
\label{mass_xi}
\end{align}

We compute $\log Z_n$ using the heat kernel representation.
The (massless) heat kernel $K_n$  for the Laplacian $\Delta_n$ 
on $n$-sheeted space $\mathcal{M}_n$ is defined as 
\begin{align}
K_n(x, x';s) = \Bra{x} e^{\Delta_n s} \Ket{x'} \,,   \quad 
\tr K_n(s) = \int_{\mathcal{M}_n} \!\rmd^d x \sqrt{g} \, K_n(x, x;s) \,. 
\end{align} 
Using the heat kernel, 
$\log Z_n$ is written as 
\begin{align}
\log Z_n = \frac12 \int^\infty_{\delta^2} \frac{\rmd s}{s} \tr K_n(s) e^{-m^2 s}\,, 
\end{align}
where $\delta$ is introduced as a UV cutoff.

In the original Euclidean AdS space (that is the case of $n=1$), 
since it is maximally symmetric, 
the heat kernel $K_1(x, x'; s )$ depends on $x$ and $x'$  only through the geodesic distance $X(x,x') $ between the two points 
(see, e.g.,  \cite{Camporesi:1990wm, Mann:1996ze, Giombi:2008vd, Fukuma:2013mx}). 
The geodesic distance can be written as 
\begin{align}
X(x,x') = \mathrm{arccosh} Z(x,x') \,,   
\end{align}
where $Z(x,x')$ is an invariant quantity under EAdS isometry, 
which is defined as a scalar product using embedding coordinates\footnote{
$X^M(x)$ satisfy $\eta_{MN} X^M(x) X^N(x) = -\ell_\text{AdS}^2 = -1$. } 
 $X^M(x)$ to flat space $\mathrm{R}^{1,d}$
\begin{align}
 Z(x,x') = - \eta_{MN} X^M(x) X^N (x') \,. 
\end{align}
We thus write the heat kernel as $K_1(X;s)$. 
If two points $x$ and $x'$ are different only in $\tau_E$-direction of the coordinates \eqref{ds_topobh} 
as $x=(\rho, \tau_E, v, \Omega_i)$ and  
$x'=(\rho, \tau_E', v, \Omega_i)$, 
the invariant $Z(x,x')$ depends only on $\rho$ and $\tau_E- \tau_E'$ as follows: 
\begin{align}
 Z(x,x') = 1+ 2 \rho^2 \sin^2 \frac{\tau_E- \tau_E'}{2} \,.
\end{align}

From the heat kernel $K_1(X,s)$ on EAdS, 
we can evaluate the heat kernel $K_n(x, x'; s )$ on $n$-sheeted space 
$\mathcal{M}_n$ by the Sommerfeld formula \cite{Sommerfeld:1897}, 
(see also \cite{Dowker:1977zj, Mann:1996ze, Solodukhin:2011gn}), 
\begin{align}
K_n(x,x'; s )= K_1(X(\tau_E,\tau_E') ;s) 
+ \frac{i}{4\pi n} \int_\Gamma \rmd w \cot\frac{w}{2 n}  K_1(X(\tau_E+w,\tau_E') ;s) \,, 
\label{som_formula}
\end{align}
where the contour $\Gamma$ consists of two lines: One goes from $(-\pi +i \infty)$ to $(-\pi -i \infty)$ 
intersecting the real axis between the poles $-2 \pi n$ and 0 of $\cot\frac{w}{2 n} $,  
and another goes from $(\pi -i \infty)$ to $(\pi +i \infty)$ 
intersecting the real axis between the poles 0 and $2 \pi n$. 
Using the formula \eqref{som_formula}, $\tr K_n(s)$  is computed as 
\begin{align}
 \tr K_n(s) &= \int^{2\pi n}_0 \rmd \tau_E \int_{H^{d-2}} \rmd V_{d-2} 
 \int^\infty_0 \! \rmd \rho \,  \rho (1+\rho^2)^{\frac{d-3}{2}} K_n(x,x; s ) \nn\\ 
 &= n \, \tr K_1(s) +\frac{i}{2}  \mathcal{A}^{(d-2)}(\Sigma)   \int^\infty_0 \! \rmd \rho \,  \rho (1+\rho^2)^{\frac{d-3}{2}}  
\int_\Gamma \rmd w \cot\frac{w}{2 n}  K_1(X ;s)\,, 
\label{tr_k_n_som}
\end{align}
where $\cosh X = 1+ 2 \rho^2 \sin^2 (w/2)$. 
Since the first term in \eqref{tr_k_n_som} is canceled in the combination $\tr K_n -n\, \tr K_1$, 
the R\'enyi entropy $S_n$ is represented as 
\begin{align}
S_n 
&= \frac{1}{1-n} (\log Z_n- n \log Z_1) \nn\\ 
&= \mathcal{A}^{(d-2)}(\Sigma)  \frac{1}{2(1-n)} \frac{i}{2}  \int^\infty_{\delta^2} \frac{\rmd s}{s} e^{-m^2 s} 
\int^\infty_0 \! \rmd \rho \,  \rho (1+\rho^2)^{\frac{d-3}{2}}  
\int_\Gamma \rmd w \cot\frac{w}{2 n}  K_1(X ;s)
\,. 
\label{renyi_int}
\end{align}  

The integrand of \eqref{renyi_int} is singular at $s=0$, and the leading singularity is 
evaluated as   
\begin{align}
 \int^\infty_0 \! \rmd \rho \,  \rho (1+\rho^2)^{\frac{d-3}{2}}  
\int_\Gamma \rmd w \cot\frac{w}{2 n}  K_1(X ;s) 
&\sim \int_\Gamma \rmd w \cot\frac{w}{2 n} \, \frac{1}{\sin^2 \frac{w}{2}} \frac{1}{8 \pi (4\pi s)^{\frac{d-2}{2}}} \nn\\
&= i \frac{n^2-1}{6n}  \frac{1}{ (4\pi s)^{\frac{d-2}{2}}} \,. 
\end{align}
Therefore, the R\'enyi entropy follows the bulk area-law  
\begin{align}
S_n = \frac{1+n}{12 n (d-2) (4 \pi)^{\frac{d-2}{2}}} \, 
\frac{\mathcal{A}^{(d-2)}(\Sigma)}{\delta^{d-2}} +\mathcal{O}(\delta^{-d+4})\,. 
\label{renyi_area-law}
\end{align}
In particular, the entanglement entropy behaves as 
\begin{align}
S_1 = \frac{1}{6 (d-2) (4 \pi)^{\frac{d-2}{2}}} \, 
\frac{\mathcal{A}^{(d-2)}(\Sigma)}{\delta^{d-2}} +\mathcal{O}(\delta^{-d+4}) \,. 
\label{area-law}
\end{align}
The area-law of entanglement entropy should hold 
because we consider entanglement entropy for the ground state of a local QFT. 
As expected, 
the behavior of the leading singularity \eqref{area-law}  is the same as in Minkowski space and other curved backgrounds (e.g. \cite{Solodukhin:2011gn}). 
We will see that the subleading parts also follow the bulk  area-law.

\section{Entanglement entropy in AdS}
\label{sec_eeads}
The leading singular term of entanglement entropy \eqref{renyi_area-law} or \eqref{area-law} depends on the definition of the  UV cutoff $\delta$.  
In odd dimensions, entanglement entropy is expected to have the form  
\begin{align}
S_1 \sim \frac{\alpha_{d-2}}{\delta^{d-2}} +  \frac{\alpha_{d-4}}{\delta^{d-4}} + \ldots + \alpha_0 . 
\end{align}
The last term is a universal term which is finite in the limit $\delta\to 0$. 
In this section, we compute the subleading terms including  the universal term by using the explicit form of the heat kernel $K_1$ on 
odd-dimensional 
Euclidean AdS space.  

We represent the heat kernel on $d$-dimensional Euclidean AdS space as $K^{(d)}_1(X;s)$. 
It is computed in \cite{Davies1988, Camporesi:1990wm, Grigor'yan} 
and takes the form\footnote{
The heat kernels are those for the Dirichlet boundary conditions, 
that is,  we sum up only normalizable modes for $m=0$ 
(see, e.g., \cite{Camporesi:1991nw}.)
} 
\begin{align}
K^{(d)}_1(X;s) = 
\left\{ \begin{array}{ll}
 (\frac{-1}{2\pi})^\frac{d-1}{2} \frac{1}{(4 \pi s)^\frac12} (\frac{1}{\sinh X} \frac{\partial}{\partial X})^\frac{d-1}{2}\,  
 e^{-(\frac{d-1}{2})^2 s - \frac{X^2}{4s}} 
 & (d:\text{odd}) \\
 (\frac{-1}{2\pi})^\frac{d-2}{2} \frac{\sqrt2}{(4 \pi s)^\frac32}  e^{-(\frac{d-1}{2})^2 s}
  (\frac{1}{\sinh X} \frac{\partial}{\partial X})^\frac{d-2}{2} 
\int^\infty_X \rmd X' \frac{X' e^{- \frac{X^{\prime 2}}{4s}}}{\sqrt{\cosh X' -\cosh X}}
 & (d:\text{even}) \\
\end{array} \right. \,.  
\label{heat_kernel_general_d}
\end{align} 
In particular for $d=3$, $K^{(3)}_1$ is 
\begin{align}
K^{(3)}_1 (X;s) = \frac{1}{(4 \pi s)^\frac32 } \frac{X}{\sinh X} \,  
 e^{- s - \frac{X^2}{4s}} \,, 
 \label{Kin3}
\end{align}
which is also  computed in \cite{Mann:1996ze, Giombi:2008vd, David:2009xg}. 

From the form \eqref{heat_kernel_general_d}, 
one can find that the heat kernels satisfy the recurrence relation 
\cite{Grigor'yan} 
\begin{align}
K^{(d+2)}_1(X;s) =-\frac{e^{-d s}}{2 \pi \sinh X}  \frac{\partial}{\partial X} K^{(d)}_1(X;s) \,. 
\label{rec_rel}
\end{align}

\subsection{Entanglement entropy in AdS$_3$}
We first consider three-dimensional AdS space. 
Using the heat kernel $K^{(3)}_1$ in \eqref{Kin3} 
and the Sommerfeld formula \eqref{som_formula}, 
we can derive the heat kernel $K^{(3)}_n$ on $n$-sheeted space $\mathcal{M}_n$. 
Here, we compute $Z_n$ in the case of $n=1/N$ ($N \in \mathbb{N}$) 
in order to confirm the Sommerfeld formula \eqref{som_formula}.   

In the case of $n=1/N$, 
$Z_{1/N}$ represents the partition function 
on the orbifold EAdS$_3 /\mathbf{Z}_N$ 
as in \cite{Nishioka:2006gr, Miyagawa:2015sql},  
and 
the trace of the heat kernel on the orbifold is easily obtained by the method of images 
\begin{align}
\tr K^{(3)}_{1/N}(s) 
&=\int^{\infty}_0 \!\rmd \rho \,\rho 
\int^{2\pi/N}_0 \!\!\!\!\! \rmd \tau_E \int^\infty_{-\infty} \!\!\!\rmd v
\sum_{j=0}^{N-1} 
\Bra{\rho, \tau_E + 2\pi j/N ,v} e^{\Delta s} \Ket{\rho, \tau_E ,v} \,. 
\label{trK3}
\end{align}
Since the heat kernel on AdS$_3$ is given by \eqref{Kin3}, 
we have 
\begin{align}
\Bra{\rho, \tau_E + w ,v} e^{\Delta s} \Ket{\rho, \tau_E ,v} 
= \frac{1}{(4 \pi s)^\frac32 } \frac{X(\rho, w)}{\sinh X(\rho, w)} \,  
 e^{- s - \frac{X(\rho, w)^2}{4s}} \,, 
\end{align}
where $\cosh X(\rho, w) = 1+ 2 \rho^2 \sin^2 (w/2)$. 
Thus, eq.\eqref{trK3} becomes 
\begin{align}
\tr K^{(3)}_{1/N}(s) & = 
\frac{1}{N} \tr K^{(3)}_{1}(s) \nn\\ 
&+ \frac{2\pi}{N} \mathcal{A}^{(1)}(\Sigma) 
\sum_{j=1}^{N-1} \frac{1}{4\sin^2 (\pi j/N)} 
\frac{e^{- s}}{(4 \pi s)^\frac32 }
\int^{\infty}_0 \! \rmd X X \,  
 e^{- \frac{X^2}{4s}} \,. 
\end{align}
Performing the integral as 
\begin{align}
\int^{\infty}_0 \! \rmd X X \,  
 e^{- \frac{X^2}{4s}} 
 = 2s \,, 
\end{align}
and applying the formula 
\begin{align}
\sum_{j=1}^{N-1} \frac{1}{\sin^2 (\pi j/N)} 
=\frac{N^2-1}{3}\,, 
\end{align}
we obtain 
\begin{align}
\tr K^{(3)}_{1/N}(s) -\frac{1}{N}\tr K^{(3)}_{1}(s)
&= \mathcal{A}^{(1)}(\Sigma) 
\frac{N^2-1}{12 N} 
\frac{ e^{- s}}{(4 \pi s)^\frac12 } 
 = \mathcal{A}^{(1)}(\Sigma) 
\frac{1-N^{-2}}{12 N^{-1}} 
\frac{ e^{- s}}{(4 \pi s)^\frac12 } \,. 
\label{trk3_1/N}
\end{align}
This also can be obtained from the formula \eqref{tr_k_n_som} for general $n$ 
as 
\begin{align}
 \tr K^{(3)}_n(s) -n \, \tr K^{(3)}_1(s)
 &=  \frac{i}{2}  \mathcal{A}^{(1)}(\Sigma) 
 \int^\infty_0 \! \rmd \rho \,  \rho 
\int_\Gamma \rmd w \cot\frac{w}{2 n}  \frac{1}{(4 \pi s)^\frac32 } \frac{X(w)}{\sinh X(w)} \,  
 e^{- s - \frac{X(w)^2}{4s}}\, \nn \\
 &= \frac{i}{2}  \mathcal{A}^{(1)}(\Sigma) \frac{e^{-s}}{4(4 \pi s)^\frac32 }
  \int^\infty_0 \! \rmd X X e^{- \frac{X^2}{4s}}
 \int_\Gamma \rmd w \cot\frac{w}{2 n}  \frac{1}{\sin^2 \frac{w}{2}} \,  \nn\\ 
 &= \mathcal{A}^{(1)}(\Sigma) 
 \frac{1-n^{2}}{12 n} 
 \frac{ e^{- s}}{(4 \pi s)^\frac12 } \,. 
\end{align}

Therefore, the R\'enyi entropy is computed as  
\begin{align}
S_n &= \frac{1}{1-n}  \frac12 \int^\infty_{\delta^2} \frac{\rmd s}{s}
 (\tr K^{(3)}_n(s) -n \, \tr K^{(3)}_1(s)) e^{-m^2 s} \nn \\
 &=\mathcal{A}^{(1)}(\Sigma) 
  \frac{1+n}{24 n} \int^\infty_{\delta^2} \frac{\rmd s}{s}
  \frac{ e^{- (m^2+1) s}}{(4 \pi s)^\frac12 } \nn\\
  &=\mathcal{A}^{(1)}(\Sigma) 
    \frac{1+n}{12 n (4 \pi)^\frac12} (\delta^{-1}-\sqrt{\pi} \nu ) \,, 
\end{align}
where we omit $\mathcal{O}(\delta)$ terms, and 
$\nu$ is defined in \eqref{def_nu}. 
We thus find that the R\'enyi entropy monotonically decreases with respect to $n$:\footnote{
The universal part monotonically increases with respect to $n$ as in \eqref{d3_renyi_univ}.} 
\begin{align}
S_n &=\Bigl(1+\frac{1}{n}\Bigr) S_\infty \,, \quad 
S_\infty = 
    \frac{\mathcal{A}^{(1)}(\Sigma)}{12 (4 \pi)^\frac12} (\delta^{-1}-\sqrt{\pi} \nu ) \,. 
\end{align}
The entanglement entropy $S_1$ is especially given by 
\begin{align}
S_1=
    \frac{\mathcal{A}^{(1)}(\Sigma) }{6  (4 \pi)^\frac12} (\delta^{-1}-\sqrt{\pi} \nu ) \,. 
\end{align}
The universal term independent of $\delta$ in $S_n$ is 
\begin{align}
-\frac{\mathcal{A}^{(1)}(\Sigma)}{24} \Bigl(1+\frac{1}{n}\Bigr)\nu \,. 
\label{d3_renyi_univ}
\end{align}
In the case of $n=1$, 
the coefficient of area $\mathcal{A}^{(1)}(\Sigma)$ is the same as in \cite{Miyagawa:2015sql}. 

In order to see the universal contribution, 
we further need to introduce the bulk IR cutoff $\epsilon$.     
Here we introduce a cutoff surface at $z=\epsilon$ 
in the Poincar\'e coordinates \eqref{poincare}.  
The area is given in \eqref{area_finite}:  
\begin{align}
 \mathcal{A}^{(1)}(\Sigma) =2 \int^1_{\epsilon/r_0} \rmd y \frac{1}{y\sqrt{1-y^2}} 
 = 2 \log \frac{2r_0}{\epsilon} +\mathcal{O}(\epsilon/r_0) \,. 
\end{align}
The universal contribution to the R\'enyi entropy $S_n$ is the term including $\log \epsilon^{-1}$ 
and is given by 
\begin{align}
S_n^{univ} = -\frac{1}{12} \Bigl(1+\frac{1}{n}\Bigr)\nu \log \frac{2 r_0}{\epsilon}  \,. 
\end{align}

\subsection{Higher dimensions}
For general dimensions, we can compute the R\'enyi entropy $S_n$ by using \eqref{renyi_int}. 
Changing the variable of integration from $\rho$ to $X=\mathrm{arccosh}(1+2\rho^2 \sin^2 (w/2))$, 
we obtain the expression 
\begin{align}
S_n 
&=   \frac{i\,\mathcal{A}^{(d-2)}(\Sigma)}{16(1-n)} 
 \int^\infty_{\delta^2} \! \frac{\rmd s}{s} e^{-m^2 s} 
\int_\Gamma \!\rmd w \frac{\cot\frac{w}{2 n}}{\sin^2 \frac{w}{2}} \,
I^{(d)} (w,s) 
\,,  
\label{gen_int_renyi}
\end{align}
where 
\begin{align}
I^{(d)}(w,s) \equiv \int^\infty_0 \! \rmd X \, \sinh X  \Bigl(1+\frac{\cosh X-1}{2\sin^2 \frac{w}{2}}\Bigr)^{\frac{d-3}{2}}  
 K^{(d)}_1(X ;s) \,. 
  \label{int_I^d}
\end{align}
From eq.\eqref{rec_rel}, 
one can find a relation 
\begin{align}
I^{(d)}(w,s)=\frac{e^{-(d-2)s}}{2 \pi} \Bigl(
 K^{(d-2)}_1(0 ;s) + \frac{d-3}{4 \sin^2 \frac{w}{2}} I^{(d-2)}(w,s) 
 \Bigr)\,.  
 \label{rel_I^d}
\end{align}
We can obtain  the expressions of $I^{(d)}$ recursively from this equation. 
We give the expressions for odd $d\leq 11$ in Appendix~\ref{appendixA}. 
Using the expressions of $I^{(d)}$, 
we next compute 
\begin{align}
\int_\Gamma \!\rmd w \frac{\cot\frac{w}{2 n}}{\sin^2 \frac{w}{2}} \,
I^{(d)} (w,s)  \,. 
\end{align}
The results are also given in Appendix~\ref{appendixA}. 

Finally performing the $s$-integral in \eqref{gen_int_renyi}, 
we obtain the R\'enyi entropy $S_n$. 
They are functions of $\nu$ which is defined in \eqref{def_nu}. 
We write the entropy $S_n$ for odd $d$-dimensional AdS as $S_n^{(d)}(\nu)$, 
which takes the following form:   
\begin{align}
S_n^{(d)} (\nu) = 
\frac{\mathcal{A}^{(d-2)}(\Sigma)}{12 (4\pi)^\frac{d-2}{2}} \Bigl(1+\frac{1}{n}\Bigr) 
\Bigl(\sum_{k=0}^{\frac{d-3}{2}} g^{(d)}_{d-2-2k}(n) \delta^{-(d-2-2k)} +g^{(d)}_0 (n) \Bigr) \,,  
\end{align}
where the expressions of $g^{(d)}_{d-2-2k}(n)$ are given in Appendix~\ref{appendixA}.   
We thus find that  not only for the leading term but also the subleading terms follow the bulk area-law. 
This is due to the facts that AdS space is maximally symmetric 
and the extrinsic curvature of the surface $\Sigma$ vanishes because $\Sigma$ is a Killing horizon.  

$g^{(d)}_0(n)$ represents a universal contribution which is finite in the limit $\delta\to 0$, 
and is given as follows 
when we omit $\mathcal{O}(\delta)$ terms: 
\begin{align}
 g^{(3)}_0(n) &= - \sqrt\pi \nu 
\,, \label{g30}
\\
 g^{(5)}_0(n) &= \frac{2}{3} \sqrt{\pi } \nu ^3-\frac{11 n^2 +1 }{15 n^2}  \sqrt{\pi } \nu   
\,,  \\
 g^{(7)}_0(n) &=-\frac{4}{15} \sqrt{\pi } \nu ^5+\frac{4  \left(16 n^2+1\right)}{45 n^2} \sqrt{\pi } \nu ^3 
-\frac{2   \left(191 n^4+23n^2+2\right)}{315 n^4} \sqrt{\pi } \nu 
\,, \\
g^{(9)}_0(n) &= \frac{8}{105} \sqrt{\pi } \nu ^7-\frac{4 \left(21 n^2+1\right)}{75 n^2}  \sqrt{\pi } \nu ^5
+\frac{40 \left(162 n^4+15n^2+1\right)}{1575 n^4} \sqrt{\pi } \nu ^3  \nn\\ 
&\quad -\frac{2 \left(11 n^2+1\right) \left(227 n^4+10 n^2+3\right)}{1575 n^6} \sqrt{\pi } \nu  
\,,\\
g^{(11)}_0 (n) &= -\frac{16}{945} \sqrt{\pi } \nu ^9 
+\frac{32  \left(26 n^2+1\right)}{1575 n^2} \sqrt{\pi } \nu ^7
-\frac{16  \left(492n^4+37 n^2+2\right)}{1575 n^4}  \sqrt{\pi } \nu ^5
\nn\\ 
&\quad +\frac{16  \left(4608 n^6+508 n^4+53 n^2+3\right)}{4725 n^6} \sqrt{\pi } \nu ^3
\nn\\
&\quad -\frac{8  \left(14797 n^8+2125 n^6+321 n^4+35 n^2+2\right)}{10395 n^8} \sqrt{\pi } \nu \,. 
\label{g110}
\end{align}
Note that they are all odd functions with respect to $\nu$ and 
negative in the range $0<\nu<1$ for arbitrary $n>0$. 
In particular, setting $n=1$, we find that $g^{(d)}_0(1)$ is given by 
\begin{align}
g^{(d)}_0(1) = - \frac{3 \sqrt{\pi}\,2^\frac{d-1}{2}}{(d-1)\,d!!}
\Bigl[\nu \prod_{k=1}^{\frac{d-1}{2}}(k^2-\nu^2) 
- d \int^\nu_0  \rmd y\,  \prod_{k=0}^{\frac{d-3}{2}}(k^2-y^2) 
\Bigr] \,.
\label{gd01}
\end{align} 

As in three-dimensions, 
we introduce the bulk IR cutoff at $z=\epsilon$. 
The area $\mathcal{A}^{(d-2)}(\Sigma)$ given in \eqref{area_finite} 
then takes the form \cite{Ryu:2006bv, Ryu:2006ef} 
\begin{align}
\mathcal{A}^{(d-2)}(\Sigma)/\Omega_{d-3} &= 
\sum^{\frac{d-3}{2}}_{k=1}p_{2k-1} \Bigl(\frac{r_0}{\epsilon}\Bigr)^{d-2k-1}
+ \frac{(-1)^{\frac{d-3}{2}}(d-4)!!}{(d-3)!!} \log\frac{r_0}{\epsilon}
+ \mathcal{O}(1) \,. 
\label{area_exp}
\\ 
 p_1&= (d-1)^{-1} \,,  \ldots \,. 
\end{align}
The term with $\log\epsilon^{-1}$ gives the universal contributions of the R\'enyi entropy $S_n$ as 
\begin{align}
S_n^{(d) univ} (\nu) &= 
\frac{\Omega_{d-3}}{12 (4\pi)^\frac{d-2}{2}} \Bigl(1+\frac{1}{n}\Bigr) 
  \frac{(-1)^{\frac{d-3}{2}}(d-4)!!}{(d-3)!!} g^{(d)}_0 (n) 
 \log\frac{r_0}{\epsilon} \,, \nn\\  
 &= \frac{(-1)^{\frac{d-3}{2}}}{3 \sqrt{\pi}\,2^\frac{d+1}{2} (d-3)!!} 
 \Bigl(1+\frac{1}{n}\Bigr) 
    g^{(d)}_0 (n) 
  \log\frac{r_0}{\epsilon} \,.  
\end{align} 
Note that 
it vanishes when $\nu=0$, that is the case where 
the Breitenlohner-Freedman bound \cite{Breitenlohner:1982jf} is saturated. 
Using \eqref{gd01}, the universal part of the entanglement entropy can be 
written as  
\begin{align}
S_1^{(d) univ} (\nu) &= 
\frac{(-1)^{\frac{d-1}{2}}}{d!} 
\Bigl[\nu \prod_{k=1}^{\frac{d-1}{2}}(k^2-\nu^2) 
- d \int^\nu_0  \rmd y\,  \prod_{k=0}^{\frac{d-3}{2}}(k^2-y^2) 
\Bigr] 
  \log\frac{r_0}{\epsilon} \,. 
\end{align}

\section{One-loop corrections to holographic entanglement entropy}
\label{sec_1loop}

\subsection{Contributions from a scalar field}
According to the FLM proposal, the 1-loop corrections 
to holographic entanglement entropy is given by \eqref{FLM}. 
We evaluate the contributions from a scalar field. 
In this section we restore a AdS radius $\ell_\text{AdS}$ which has been set to $\ell_\text{AdS}=1$, 
but use the symbol $\mathcal{A}^{(d-2)}(\Sigma)$ 
to represent the minimal area with $\ell_\text{AdS}=1$, 
that is, the  integral \eqref{area}. 

We have found that the bulk entanglement entropy of a scalar field 
is 
\begin{align}
S^{(d)}_\text{bulk} &=  
\frac{\mathcal{A}^{(d-2)}(\Sigma)}{6 (4\pi)^\frac{d-2}{2}} 
g^{(d)}_0 (1)  \nn \\
&= - \frac{\mathcal{A}^{(d-2)}(\Sigma)}{2 (2\pi)^\frac{d-3}{2} (d-1)\,d!!}
\Bigl[\nu \prod_{k=1}^{\frac{d-1}{2}}(k^2-\nu^2) 
- d \int^\nu_0  \rmd y\,  \prod_{k=0}^{\frac{d-3}{2}}(k^2-y^2) 
\Bigr] \,,  
\label{s_bulk}
\end{align}
where we ignore the bulk UV divergent terms.

We next evaluate the change of minimal area due to the back reaction 
from a quantum expectation value of the energy momentum tensor 
$T_{\mu\nu}$ of the scalar field. 
The expectation value is evaluated in \cite{Caldarelli:1998wk}, 
and is given for odd-dimensional AdS$_d$ as follows 
\begin{align}
\braket{T_{\mu\nu}}_d = - g_{\mu\nu}  
\frac{(m_0 \, \ell_\text{AdS})^2}{2 (2\pi)^\frac{d-1}{2} d!!\, 
\nu\, \ell_\text{AdS}^d} 
\prod_{k=0}^{\frac{d-3}{2}}(k^2-\nu^2) 
\equiv \lambda g_{\mu\nu} \,,  
\label{em_vev}
\end{align}
where we also omit the bulk UV divergences and separate the mass $m_0$ from curvature coupling $\xi$ 
as \eqref{mass_xi}, so $\nu$ is 
\begin{align}
\nu = \sqrt{(m_0 \, \ell_\text{AdS})^2 -d(d-1) \xi + \biggl(\frac{d-1}{2}\biggr)^2} \,. 
\label{nu_xi}
\end{align}
As discussed in \cite{Miyagawa:2015sql},   
we also have a pure AdS solution of the Einstein equation, 
\begin{align}
R_{\mu\nu}-\frac12 R g_{\mu\nu} +\Lambda g_{\mu\nu}
= 8\pi G_{N} \braket{T_{\mu\nu}}_d \,,  
\end{align}
if the energy momentum tensor has a expectation value 
with the form $\braket{T_{\mu\nu}}_d = \lambda g_{\mu\nu}$.  
Since the cosmological constant $\Lambda$ is related to  a AdS radius 
as 
\begin{align}
\Lambda = -\frac{(d-1)(d-2)}{2 \ell_\text{AdS}^2}, 
\end{align}
a shift of the AdS radius due to the back reaction is given by   
\begin{align}
\delta \ell_\text{AdS} = 
- \frac{8\pi G_N\ell_\text{AdS}^3}{(d-1)(d-2)} \lambda 
+ \mathcal{O}(G_N^2)   
\end{align}
as in \cite{Miyagawa:2015sql}. 
Noting that the classical holographic entanglement entropy is 
\begin{align}
\frac{\text{area}(\Sigma)}{4 G_N} = 
\frac{\ell_\text{AdS}^{d-2}\,\mathcal{A}^{(d-2)}(\Sigma)}{4\, G_N} \,, 
\end{align}
we find that the second term in \eqref{FLM} takes the form   
\begin{align}
\delta \frac{\text{area}(\Sigma)}{4 G_N} &= 
- \frac{2\pi \, \ell_\text{AdS}^{d}\,\mathcal{A}^{(d-2)}(\Sigma)}{d-1}
 \lambda \\ 
 &= \frac{ \mathcal{A}^{(d-2)}(\Sigma) (m_0 \, \ell_\text{AdS})^2 
}{2 (2\pi)^\frac{d-3}{2} d!!\,(d-1)\, 
 \nu} 
 \prod_{k=0}^{\frac{d-3}{2}}(k^2-\nu^2) \,. 
 \label{shift_minarea}
\end{align}
In the last line we have used the explicit expression, \eqref{em_vev}, of $\lambda$. 

The third term $\delta S_\text{Wald}$ in \eqref{FLM} comes from 
a curvature coupling term $-\xi R \phi^2/2$. 
If $\phi^2$ has a expectation value, the term plays a role of  
the Einstein-Hilbert term, 
which also contributes to the Bekenstein-Hawking entropy or the Wald entropy 
as 
\begin{align}
\delta S_\text{Wald} = -2\pi \xi \braket{\phi^2}_d \, \text{area}(\Sigma) \,. 
\end{align}
In odd-dimensional space, we have a relation \cite{Caldarelli:1998wk} 
\begin{align}
\braket{T^{\mu}_{\phantom{\mu}\mu}}_d = -m_0^2 \braket{\phi^2}_d\,. 
\end{align}
Thus, the renormalized expectation value of $\phi^2$ is 
\begin{align}
\braket{\phi^2}_d = -\frac{d}{m_0^2} \lambda \,. 
\end{align}
Therefore, we obtain 
\begin{align}
\delta S_\text{Wald} &= \frac{2 \pi d\, \ell_\text{AdS}^{d-2}\,\mathcal{A}^{(d-2)}(\Sigma)}{m_0^2} \xi\, \lambda \\
&= -\frac{ \mathcal{A}^{(d-2)}(\Sigma) \, \xi 
}{2 (2\pi)^\frac{d-3}{2} (d-2)!!\,
 \nu} 
 \prod_{k=0}^{\frac{d-3}{2}}(k^2-\nu^2) \,. 
 \label{ee_wald}
\end{align}

If we combine \eqref{shift_minarea} and \eqref{ee_wald}, 
we have 
\begin{align}
\delta \frac{\text{area}(\Sigma)}{4 G_N} + \delta S_\text{Wald} 
= \frac{\mathcal{A}^{(d-2)}(\Sigma)}{2 (2\pi)^\frac{d-3}{2} (d-1)\,d!! }
\nu \prod_{k=1}^{\frac{d-1}{2}}(k^2-\nu^2) \,.
\end{align}
Adding this to \eqref{s_bulk}, 
the first term in the square bracket of \eqref{s_bulk} cancels out. 
Thus, at $\mathcal{O} (G_N^0)$, 
the holographic entanglement entropy is given by $S_{cl}+S_q$ 
with 
\begin{align}
S_{cl} &= 
\frac{\ell_\text{AdS}^{d-2}\,\mathcal{A}^{(d-2)}(\Sigma)}{4\, G_N} \,, 
\label{scl_D}\\ 
S_q  &= \frac{\mathcal{A}^{(d-2)}(\Sigma)}
{2 (2\pi)^\frac{d-3}{2}(d-1)\,(d-2)!! }
\int^\nu_0  \rmd y\,  \prod_{k=0}^{\frac{d-3}{2}}(k^2-y^2) 
\,. 
\label{sq_D}
\end{align}

\subsection{The shift of entanglement entropy under a double trace deformation} 
We finally compute the shift of entanglement entropy under 
an RG flow triggered by a double trace deformation. 
Since the heat kernels \eqref{heat_kernel_general_d},  
which we have used so far,  
are those for the Dirichlet boundary condition, 
$1/N$ corrections to the entanglement entropy for CFT$^{(D)}$ 
are holographically given by \eqref{sq_D}. 
Due to the fact that Green's function for the Neumann boundary condition 
is formally obtained by flipping the sign of $\nu$ 
(see \cite{Gubser:2002zh,Miyagawa:2015sql}), 
1-loop corrections $S_q$ of entanglement entropy for CFT$^{(N)}$ 
is obtained as  
\begin{align}
S_q^{(N)}  &= \frac{\mathcal{A}^{(d-2)}(\Sigma)}
{2 (2\pi)^\frac{d-3}{2} (d-1)\,(d-2)!! }
\int^{-\nu}_0  \rmd y\,  \prod_{k=0}^{\frac{d-3}{2}}(k^2-y^2) 
= - S_q^{(D)} 
\,. 
\end{align}
Therefore, 
the shift of entanglement entropy is 
\begin{align}
S_q^{(N)} - S_q^{(D)}  &= -2 S_q^{(D)} 
=-\frac{\mathcal{A}^{(d-2)}(\Sigma)}
{(2\pi)^\frac{d-3}{2} (d-1)\,(d-2)!! }
\int^\nu_0  \rmd y\,  \prod_{k=0}^{\frac{d-3}{2}}(k^2-y^2) \,. 
\label{sqN-sqD}
\end{align} 
The coefficient of area $\mathcal{A}^{(d-2)}(\Sigma)$ is positive 
in the range $0<\nu<1$. 
Using the expansion of $\mathcal{A}^{(d-2)}(\Sigma)$ in \eqref{area_exp}, 
we find the shift of the universal term, 
\begin{align}
S_q^{(N)univ} - S_q^{(D)univ} = 
(-1)^{\frac{d-1}{2}}
\frac{2}{(d-1)!} \Bigl[
 \int^\nu_0  \rmd y\,  \prod_{k=0}^{\frac{d-3}{2}}(k^2-y^2) \Bigr]
 \, \log\frac{r_0}{\epsilon} \,. 
 \label{sqN-sqD_univ}
\end{align}
This is related to the shift of the A-type anomaly 
between  CFT$^{(N)}$ and  CFT$^{(D)}$ 
through \eqref{suniv_anomaly}. 
From \eqref{sqN-sqD_univ}, 
the shift of the anomaly is given by 
\begin{align}
\delta a^{\ast}_{d-1} = -\frac{1}{2 (d-1)!}
 \int^\nu_0  \rmd y\,  \prod_{k=0}^{\frac{d-3}{2}}(k^2-y^2) \,. 
\end{align}
Since the shift is always positive in the range $0<\nu<1$, 
the central charge associated with the A-type trace anomaly for the UV fixed point, CFT$^{(N)}$, 
is bigger than that for the IR fixed point, CFT$^{(D)}$.  
This is consistent with Zamolodchikov's c-theorem 
in two-dimensional CFT \cite{Zamolodchikov:1986gt}, 
Cardy's a-theorem for four-dimensional CFT 
\cite{Cardy:1988cwa,Komargodski:2011vj}, 
and the holographic c-theorem in higher dimensions \cite{Myers:2010tj}.

The ratio of \eqref{sqN-sqD} to $S_{cl}$ \eqref{scl_D} is given by 
\begin{align}
\frac{S_q^{(N)} - S_q^{(D)} }{S_{cl}} = 
-\frac{4 G_N}
{(2\pi)^\frac{d-3}{2} (d-1)\,(d-2)!!  \ell_\text{AdS}^{d-2}}
\int^\nu_0  \rmd y\,  \prod_{k=0}^{\frac{d-3}{2}}(k^2-y^2) \,. 
\label{ratio_ee}
\end{align}
On the other hand, the shift of 1-loop vacuum energy density between two boundary conditions in AdS$_d$ is computed in \cite{Gubser:2002zh,Gubser:2002vv}, 
and  takes the form 
\begin{align}
V^{(N)} - V^{(D)} = 
-\frac{1}
{(2\pi)^\frac{d-1}{2}  (d-2)!! \ell_\text{AdS}^{d}}
\int^\nu_0  \rmd y\,  \prod_{k=0}^{\frac{d-3}{2}}(k^2-y^2) \,. 
\end{align}
Therefore, we obtain a relation 
\begin{align}
\frac{S_q^{(N)} - S_q^{(D)} }{S_{cl}} =\frac{8\pi G_N \ell_\text{AdS}^{2}}{d-1}
(V^{(N)} - V^{(D)})\,. 
\label{ee_vac_en}
\end{align}
This is consistent with the relation between the central charge and 
the vacuum energy in AdS \cite{Gubser:2002zh,Miyagawa:2015sql}: 
\begin{align}
\frac{\delta a}{a} = \frac{8\pi G_N \ell_\text{AdS}^{2}}{d-1} 
(V^{(N)} - V^{(D)}) \,. 
\end{align}
Note that \eqref{ee_vac_en} holds without restricting the universal part of the entanglement entropy 
because $\epsilon$-dependence, $\mathcal{A}^{(d-2)}(\Sigma)$,  
cancels in the ratio \eqref{ee_vac_en}. 

\section{Summary}
\label{sec_sum}
In the paper, we have computed entanglement entropy for free massive scalar fields in AdS space. 
Using the replica trick, it can be computed from a thermal free energy of 
the topological black hole. 
We have evaluated the free energy by the heat kernel method. 
We have obtained analytical expressions of the R\'enyi entropy for 
odd-dimensional AdS up to $d=11$. 

Following the FLM proposal \eqref{FLM}, we have also evaluated 
1-loop corrections to holographic entanglement entropy 
contributed  from a scalar field in the bulk. 
The contributions give the leading difference of entanglement entropy 
for two CFTs related by a double trace deformation. 
The results are consistent with c-theorems and 
the results in \cite{Gubser:2002zh,Gubser:2002vv}. 
Thus, our result provides a check of the FLM proposal.  

In the evaluation of 1-loop corrections \eqref{FLM}, 
we have assumed that $S_{c.t.}$ just cancels the bulk UV divergence. 
We leave it to future work to see that the assumption is consistent with 
the renormalization in the effective gravitational action 
due to quantum matter fields (see, e.g., \cite{Solodukhin:2011gn}). 

We have not obtained explicit expressions of the R\'enyi entropy 
for even-dimensional AdS space. 
It can be obtained from the formula \eqref{renyi_int} 
and the heat kernel \eqref{heat_kernel_general_d}.  
It is interesting to compute the entanglement entropy for more general 
asymptotically AdS space. 
It is important to extend our analysis to fermions and higher spin fields. 
In particular, 1-loop computations of supergravity theories may be interesting.

\section*{Acknowledgments} 
The author would like to thank 
Tatsuma Nishioka, Noburo Shiba and Norihiro Tanahashi 
for useful discussions. 
The work is partly supported by the Grant-in-Aid for JSPS Research Fellow Grant Number JP16J01004. 
\appendix
\section{Some computations}
\label{appendixA}

In this appendix, we summarize some computations used in the paper. 

The integrations \eqref{int_I^d}, 
\begin{align}
I^{(d)}(w,s) = \int^\infty_0 \! \rmd X \, \sinh X  \Bigl(1+\frac{\cosh X-1}{2\sin^2 \frac{w}{2}}\Bigr)^{\frac{d-3}{2}}  
 K^{(d)}_1(X ;s) \,,  
\end{align}
can be computed recursively using the relation \eqref{rel_I^d}. 
For odd $d$, 
they take the forms  
\begin{align}
I^{(d)} = \frac{e^{-(\frac{d-1}{2})^2 s}}{2 \pi (4\pi s)^\frac{d-2}{2}} \tilde{I}^{(d)} 
\end{align} 
with 
\begin{align}
\tilde{I}^{(3)} &= 1 \,, \\
\tilde{I}^{(5)} &= 1 + \frac{1}{ \sin^2 \frac{w}{2}} s\,, \\
\tilde{I}^{(7)} &= 1 +\frac{2 s}{3} + \frac{2}{ \sin^2 \frac{w}{2}} s + \frac{2}{ \sin^4 \frac{w}{2}} s^2 \,, \\
\tilde{I}^{(9)} &= 1 +2 s+ \frac{16 s^2}{15} + \frac{1}{ \sin^2 \frac{w}{2}} s(3+2s) + \frac{6}{ \sin^4 \frac{w}{2}} s^2 
+ \frac{6}{ \sin^6 \frac{w}{2}} s^3\,, \\
\tilde{I}^{(11)} &= 1 +4 s+ \frac{28 s^2}{5} + \frac{96 s^3}{35}+ \frac{4}{ 15 \sin^2 \frac{w}{2}} s(15+30s+16 s^2) \nn\\
& \quad + \frac{4}{ \sin^4 \frac{w}{2}} s^2 (3+2s)
+ \frac{24}{ \sin^6 \frac{w}{2}} s^3 
+ \frac{24}{ \sin^8 \frac{w}{2}} s^4 \,. 
\end{align}

We next define $J_d(n,s)$ as 
\begin{align}
J_d(n,s) = 
\frac{3n}{4 \pi i (n^2-1)}\int_\Gamma \rmd w \frac{\cot\frac{w}{2 n}}{\sin^2 \frac{w}{2}} \tilde{I}^{(d)} \,. 
\end{align}
They have the following expressions:  
\begin{align}
J_{3} &= 1\,,\\
J_{5} &= 1+\frac{1+11n^2}{15 n^2}s  \,,\\
J_{7} &= 1+\frac{2(1+16 n^2)}{15 n^2}s  
+\frac{2(2 + 23 n^2 + 191 n^4)}{315 n^4}s^2\,,\\
J_{9} &= 1+\frac{1+21 n^2}{5 n^2}s  
+\frac{4(1 + 15 n^2 + 162 n^4)}{105 n^4}s^2 \nn\\
&\quad +\frac{2(3 + 43 n^2 + 337 n^4 + 2497 n^6)}{1575 n^6}s^3\,,\\
J_{11} &= 1+\frac{4(1+26 n^2)}{15 n^2}s  
+\frac{4(2 + 37 n^2 + 492 n^4)}{105 n^4}s^2 
 +\frac{8(3 + 53 n^2 + 508 n^4 + 4608 n^6)}{1575 n^6}s^3 \nn\\ 
&\quad +\frac{8(2 + 35 n^2 + 321 n^4 + 2125 n^6 + 14797 n^8)}{10395 n^8}s^4\,. 
\end{align}

Using $J_d(n,s)$, 
the R\'enyi entropy $S_n^{(d)}(\nu)$ in \eqref{gen_int_renyi} 
are expressed as
\begin{align}
 S_n^{(d)}(\nu) = 
 \frac{\mathcal{A}^{(d-2)}(\Sigma)}{24} \Bigl(1+\frac{1}{n}\Bigr) 
 \int^\infty_{\delta^2} \! \frac{\rmd s}{s} \frac{e^{-\nu^2 s}}{ (4\pi s)^\frac{d-2}{2}} J_d(n,s) \,, 
\end{align}
and take the following forms after the $s$-integration: 
\begin{align}
S_n^{(d)} (\nu) = 
\frac{\mathcal{A}^{(d-2)}(\Sigma)}{12 (4\pi)^\frac{d-2}{2}} \Bigl(1+\frac{1}{n}\Bigr) 
\Bigl(\sum_{k=0}^{\frac{d-3}{2}} g^{(d)}_{d-2-2k} \delta^{-(d-2-2k)} +g^{(d)}_0 \Bigr) \,.   
\end{align}
The last terms $g^{(d)}_0$ are given by \eqref{g30}-\eqref{g110}. 
The coefficients of singular terms $g^{(d)}_{d-2-2k} $ are 
\begin{align}
 g^{(3)}_{1} &= 1 \,, 
\\
 g^{(5)}_{3} &= \frac{1}{3} \,, \quad
 g^{(5)}_{1} = -\nu^2 + \frac{11 n^2 +1}{15 n^2} \,, 
\\ 
 g^{(7)}_{5} &= \frac{1}{5} \,, \quad
 g^{(7)}_{3} = -\frac{1}{3} \nu^2 + \frac{2(16 n^2 +1)}{45 n^2} \,, \\
 g^{(7)}_{1} &= \frac12 \nu^4 -\frac{2(16 n^2 +1)}{15 n^2}\nu^2 + \frac{382 n^4 + 46 n^2 +4}{315 n^4} \,,
\\ 
 g^{(9)}_{7} &= \frac{1}{7} \,, \quad
 g^{(9)}_{5} = -\frac{1}{5} \nu^2 + \frac{21 n^2 +1}{25 n^2}\,, \\
 g^{(9)}_{3} &= \frac{1}{6} \nu^4-\frac{21 n^2 +1}{15 n^2}  \nu^2 +  \frac{4(162 n^4 + 15 n^2 +1)}{315 n^4} \,, \\
 g^{(9)}_{1} &= -\frac{1}{6} \nu^6 +\frac{21 n^2 +1}{10 n^2}  \nu^4 -\frac{4(162 n^4 + 15 n^2 +1)}{105 n^4}\nu^2 
                 + \frac{2(2497n^6+337 n^4 + 43 n^2 +3)}{1575 n^6} \,,
\\ 
 g^{(11)}_{9} &= \frac{1}{9} \,, \quad 
 g^{(11)}_{7} = -\frac{1}{7}\nu ^2 +\frac{4 \left(26 n^2+1\right)}{105 n^2} \,, \\
 g^{(11)}_{5} &= \frac{1}{10}\nu ^4 -\frac{4 \left(26 n^2+1\right)}{75 n^2}\nu ^2 
                   +\frac{4 \left(492 n^4+37 n^2+2\right)}{525 n^4} \,, \\
 g^{(11)}_{3} &=-\frac{1}{18}\nu ^6+\frac{2  \left(26 n^2+1\right)}{45 n^2}\nu ^4 \nn\\
          &\quad -\frac{4 \left(492 n^4+37 n^2+2\right)}{315 n^4} \nu ^2 
          +\frac{8 \left(4608 n^6+508 n^4+53 n^2+3\right)}{4725 n^6} 
\,, \\
 g^{(11)}_{1} &= \frac{1}{24}\nu ^8-\frac{2 \left(26 n^2+1\right)}{45 n^2}\nu ^6  
+\frac{2 \left(492 n^4+37 n^2+2\right)}{105 n^4}  \nu ^4\nn\\
&\quad -\frac{8  \left(4608 n^6+508 n^4+53 n^2+3\right)}{1575 n^6} \nu ^2
+\frac{8 \left(14797 n^8+2125 n^6+321 n^4+35 n^2+2\right)}{10395 n^8} 
\,. 
\end{align}


\end{document}